\documentclass[onecolumn]{aa} % for a paper on 1 column, quand il y a de longues formules math\'ematiques 
\usepackage{graphicx}
\usepackage[T1]{fontenc}
\usepackage{amsmath}
\usepackage{graphicx}
\usepackage{txfonts}
\usepackage{natbib} % bibliographie, avec \citet et \citep
\bibpunct{(}{)}{;}{a}{}{,} % to follow the A&A style

\begin{document}

\title{Radio emissions from pulsar companions : a refutable explanation for galactic transients and fast radio bursts}

 %\subtitle{I. Overviewing the $\kappa$-mechanism}

 \author{F. Mottez \inst{1} 
 \and
 P. Zarka \inst{2}
 }

 %\offprints{S. Bonazzola}

 \institute{LUTH, Observatoire de Paris, CNRS, Universit\'e Paris Diderot,
 5 place Jules Janssen, 92190 Meudon, France.\\
 \email{fabrice.mottez@obspm.fr}
 \and
			 LESIA, Observatoire de Paris, CNRS, UPMC, Universit\'e Paris Diderot,
 5 place Jules Janssen, 92190 Meudon, France.\\
 \email{philippe.zarka@obspm.fr}
 }

 Version du~: \today 
 %\date{Received September 15, 1996; accepted March 16, 1997}

%\abstract{}{}{}{}{} 
% 5 {} token are mandatory
 
 %\abstract{} 
 %% context heading (optional)
 %% {} leave it empty if necessary 
 % {}%heading}
 \abstract
 %% context heading (optional), leave it empty if necessary
 {The six known highly dispersed fast radio bursts are attributed to extragalactic radio sources, of unknown origin but extremely energetic. We propose here a new explanation -- not requiring an extreme release of energy -- involving a body (planet, asteroid, white dwarf) orbiting an extragalactic pulsar.}
 %% aims heading (mandatory)
 {We investigate a theory of radio waves associated to such pulsar-orbiting bodies. We focus our analysis on the waves emitted from the magnetic wake of the body in the pulsar wind. After deriving their properties, we compare them with the observations of various transient radio signals in order to see if they could originate from pulsar-orbiting bodies.}
 %% methods heading (mandatory)
 {The analysis is based on the theory of Alfv\'en wings: for a body immersed in a pulsar wind, a system of two stationary Alfv\'en waves is attached to the body, provided that the wind is highly magnetized. When destabilized through plasma instabilities, Alfv\'en wings can be the locus of strong radio sources convected with the pulsar wind. {Assuming a cyclotron maser instability operating in the Alfv\'en wings}, we make predictions about the shape, frequencies and brightness of the resulting radio emissions. }
 %% results heading (mandatory)
 {Because of the beaming by relativistic aberration, the signal is seen only when the companion is perfectly aligned between its parent pulsar and the observer, as for occultations. For pulsar winds with a high Lorentz factor ($\geq 10^4$), the whole duration of the radio event does not exceed a few seconds, and it is composed of one to four peaks lasting a few milliseconds each, detectable up to distances of several Mpc. The Lorimer burst, the three isolated pulses of PSR J1928+15, and the recently detected fast radio bursts are all compatible with our model. According to it, these transient signals should repeat periodically with the companion's orbital period. }
 % {}%conclusions}
 {The search of pulsar-orbiting bodies could be an exploration theme for new- or next-generation radio telescopes.}

 \keywords{pulsar -- exoplanet -- white dwarf -- asteroid -- magnetosphere -- radio emission -- Lorimer burst -- fast radio bursts -- radio transients -- Alfv\'en wings }
 \titlerunning{Transient radio emissions from pulsar-orbiting bodies}
 \authorrunning{Mottez and Zarka}
 \maketitle
%
%________________________________________________________________ 

%\newpage

\section{Introduction}

\citet{Lorimer_2007_giant_pulse} discovered the first highly dispersed, bright radio burst in a Parkes pulsar survey. The time--frequency dispersion of this burst perfectly followed a $1/f^{-2}$ law with a dispersion measure (DM) $\sim$375 pc.cm$^{-3}$, out of which no more than 15\% could be attributed to wave path in our Galaxy, thus its origin was supposed to be extragalactic, with a source distance of the order of 500 Mpc ($z=0.12$), well outside the local group of galaxies. As a consequence of this large distance and the large burst intensity (30 Jansky -- 1 Jy = $10^{-26}$ Wm$^{-2}$Hz$^{-1}$), the inferred emitted power was extremely large, $\sim10^{33}$ J \citep{Thornton_2013}. 
The fixed--frequency duration of this burst is $\le5$ ms, with no detectable scatter--broadening, so that the natural origin of this exceptional burst was questioned. But recently, five similar burtst (albeit less intense, $0.4-1.3$ Jy) have been found in Parkes pulsar surveys \citep{Keane_2011b,Thornton_2013}, and called ``Fast Radio Bursts'' (FRB). They emanate from different regions of the sky. Their fixed--frequency durations extend from less than 1.1 ms to 5.6 ms. Interestingly, one of these bursts also exhibits an exponential scattering tail, supporting its natural origin. With DMs between $\sim$550 and 1100 pc.cm$^{-3}$, all these bursts are attributed to extragalactic radio sources, at cosmological distances in the range $1.7-3.2$ Gpc. Their physical origin is unknown but again assumed to be extremely energetic. No event such as a supernova or a gamma ray burst was observed simultaneously with an FRB.  Proposed explanations include the annihilation of a mini black hole \citep{Keane_2012}, binary white dwarf mergers \citep{Kashiyama_2013}, neutron star mergers \citep{Totani_2013}, or implosions of supra-massive neutron stars shortly after their birth \citep{Zhang_2014}. All these explanations imply that fast radio bursts are isolated events, happening only once from a given source. Other scenarii might lead to irregularly repeatable pulses, such as a giant pulse from a young pulsar with a low burst rate \citep{Keane_2012}, or rare eruptions of flaring main-sequence stars within one kpc \citep{Loeb_2014}.

We propose here a new explanation, that does not require an extreme release of energy, while being consistent with the observations. This explanation involves a body orbiting a pulsar: planet, asteroid, white dwarf or even another neutron star. Many such objects are known in our Galaxy. Most millisecond pulsars are in binary systems, the companion being in many cases a white dwarf \citep{Savonije_1987} -- possibly of low mass \citep{Bailes_2011} -- or a neutron star \citep{Deller_2013}. Triple systems containing at least a pulsar and a white dwarf have been found \citep{Ransom_2014}. Five planets, distributed within three planetary systems, are known to orbit pulsars \citep{Wolszczan_1992, Thorsett_1993, Bailes_2011}. All of them were detected by analyzing the shift of the pulsar period $P_{obs}$ in terms of the position of the barycentre of the star, which led to deduce that it was induced by the motion of orbiting planets. It is likely that other pulsar planets remain to be discovered. Finally, asteroid belts are likely to exist around pulsars and they have been invoked by several authors for explaining timing irregularities \citep{Shannon_2013}, anti-glitches \citep{Huang_2014} or burst intermittency \citep{Cordes_2008, Deneva_2009,Mottez_2013b}. In this paper we investigate the ability of these bodies immersed in a pulsar wind to produce radio waves.

\citet{Mottez_2011_AWW} showed that a pulsar-orbiting body necessarily moves in the wind of the pulsar and not in the light-cylinder, and that its interaction with the wind causes a system of strong electric currents flowing along the body and in the plasma carried by the wind. The power associated to these currents can be higher than $10^{22}$ W. We explore what happens if a fraction of this power is converted into radio-frequency electromagnetic waves. Without needing to go into the detailed physics of the electromagnetic waves production, we can make predictions about the properties of the subsequent radio waves and we study if and how this radio signal could be detected by ground based radio-telescopes.

Our analysis is based on the electrodynamic interaction of the orbiting body and the pulsar wind derived in \citet{Mottez_2011_AWO,Mottez_2011_AWW} and \citet{mottez_2012c}. This interaction is the relativistic analogue of the Io-Jupiter interaction, that has been described in terms of a unipolar inductor circuit or a pair of Alfv\'en wings (see the review by \citet{Saur_2004}) and transposed to exoplanet-star interaction by \citet{Zarka_2001,Willes_2005,Zarka_2007}. In our case, the magnetized plasma flow around the body is not corotating like in the previous papers, but is a wind as in \citet{Preusse_2006} but here strongly relativistic.
 
In section \ref{sec_wind} we analyze the system of Alfv\'en wings generated by the interaction of the body with the pulsar wind. Then, after eliminating the possibility that the radio emission could come from the vicinity of the body (section \ref{sec_sources_attached}), we study the generation and the characteristics of the radio emission produced by a wind-convected source along the Alfv\'en wings (section \ref{sec_sources_wind}). In section \ref{possible_explanation} we compare our predictions to past observations of a few remarkable radio transients. Finally in section \ref{sec_conclusion} we address specific questions raised by our explanation, and provide perspectives for observations that can confirm or refute our theory.
% In particular, we explore the possibility that remarkable and still unexplained radio transients such as the above-mentioned fast radio bursts be interpreted in terms of these radio signals.
%We consider the two main families of pulsars. The "standard pulsars" have a period $P_* \sim 1$ s, and $\dot P \sim 10^{-14}-10^{-12}$. The second family, to which belong PSR 1257+12 (3 planets) and PSR 1620-26 (1 planet) and PSR J1719-1438 are characterized by a period $P_* \sim 10$ ms and $\dot P \sim 10^{-20}-10^{-17}$. This is the family of the "millisecond pulsars". In the following pages, we will refer to these two families of pulsars for numerical applications.

\section{Interaction with the pulsar wind} \label{sec_wind}

A body orbiting a pulsar with a period exceeding a few minutes is immersed in the pulsar wind, and the interaction between the pulsar and its companion is ruled by electrodynamic processes. The distance of the known pulsar companions is typically several hundreds of light cylinder radii ($r_{LC}$).

Among the many pulsar wind models, those of \citet{Goldreich_Julian_1969} and \citet{Michel_1969}, or more recent ones \citep{Bucciantini_2006}, show that for a pulsar with an aligned magnetic field, beyond a distance of hundreds of $r_{LC}$ from the star, the plasma wind flow is radial ($v_0 \sim v_0^r$
%, i.e. along the direction $r$ in spherical coordinates
), the magnetic field energy dominates the plasma kinetic energy ($B^2 \gg \mu_0 \rho \gamma c^2$ -- the wind is said to be "Poynting flux dominated"), and the plasma velocity is highly relativistic. 
The Lorentz factor is unknown but expected to be in the range $\gamma \sim 10^1-10^7$. The magnetic field is quasi azimuthal ($B \sim B_{\phi} \gg B_{poloidal}$). The flow and Alfv\'en velocities ($v_0$ and $V_A$) are very close to the speed of light and, if the above quoted models are correct, the wind is sub-Alfv\'enic: $v_0 < V_A$ \citep{Mottez_2011_AWW}.

\subsection{Alfv\'en wings} \label{sec_AW}

Because the fast magnetosonic waves are even faster than $V_A$, the wind is also slower than the fast magnetosonic waves, and there is thus no MHD shock ahead of an obstacle in the wind. The pulsar-orbiting body is thus in direct contact with the wind. Following the idea developed in \citet{Neubauer_1980} and applied by \citet{Mottez_2011_AWW} in the context of pulsar winds, a pair of electric current circuits are induced by the body--wind interaction, carried by two stationary Alfv\'enic structures called Alfv\'en wings. These wings are attached to the body on one side, and extend far into space into the flowing plasma on the other side. %The configuration of these electric currents is recalled in Fig. \ref{fig_inducteur_unipolaire}.

The azimuthal magnetic field of the pulsar wind is
\begin{equation} \label{eq_b0phi}
B_0^\phi=B_0^r \frac{v_0^\phi -\Omega_* r}{v_0^r} \sim -\frac{B_0^r \Omega_* r}{c} = -\frac{B_*^r \Omega_* R_*^2}{rc}
\end{equation}
where $B_*^r$, $\Omega_*$ and $R_*$ are respectively the star surface radial magnetic field, its rotation circular frequency and radius.
The convection electric field $E_0 = v_0^r B_0^\phi = B_0^r \Omega_* r$, directed perpendicularly to the wind flow and to the magnetic field
is the engine of the Alfv\'en wings. It induces a potential drop $U=2 R_b E_0$ along the body of radius $R_b$, that is the cause of a current flowing along the planet and in the surrounding plasma.

\begin{figure}
\resizebox{\hsize}{!}{\includegraphics{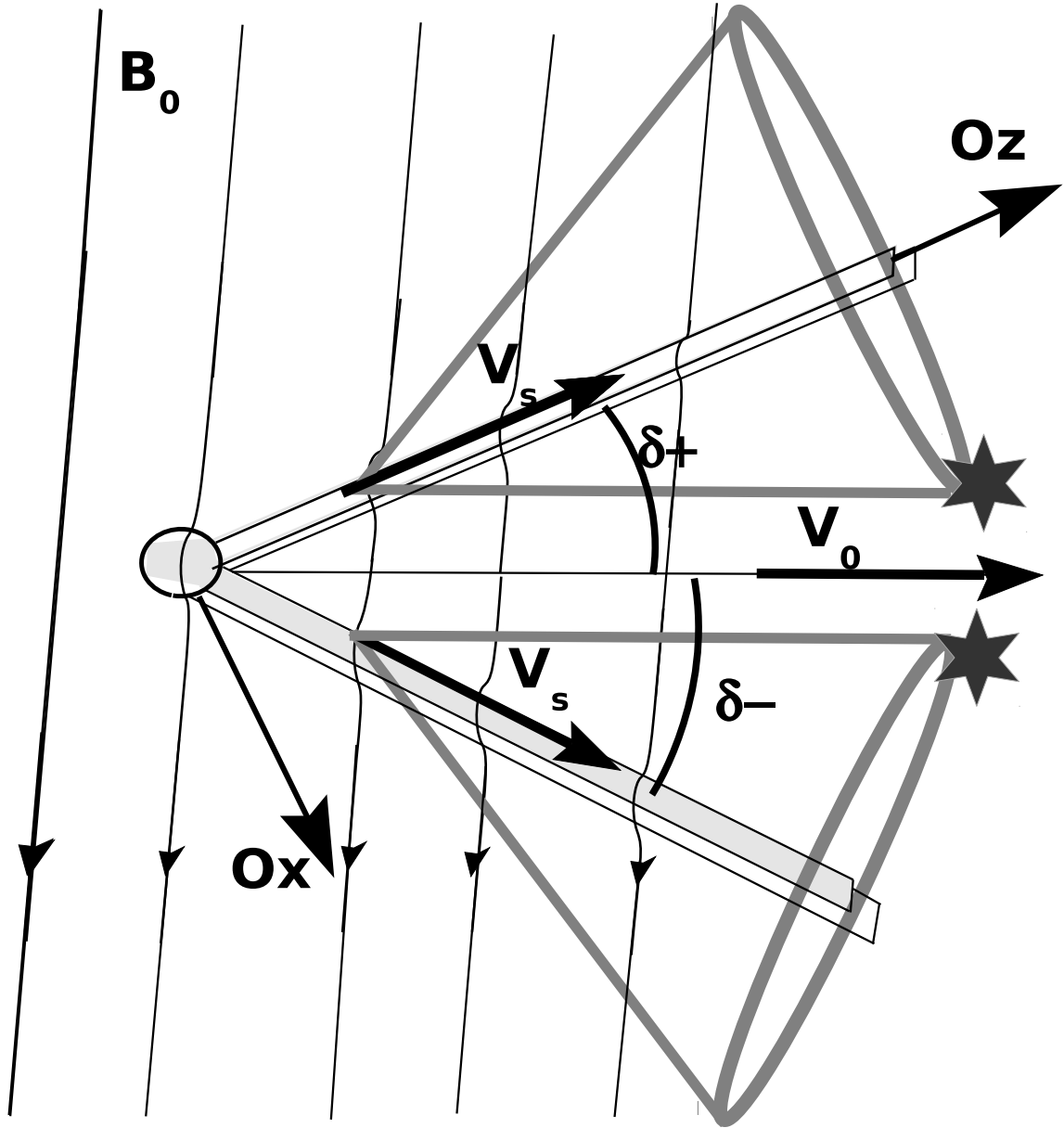}}
\caption{Alfv\'enic wake of the pulsar's companion body seen from above the equatorial plane. 
The velocity of the source regions $\bf{V}_s$ is assumed to be along the Alfv\'en wings.
The thick grey lines represent the cones in which emitted radio waves are focussed. We call them the angular emission patterns. 
The observer's frame related to the the upper wing on the figure (making an angle $\delta_+$ with the wind) is indicated by the {\bf Ox} and {\bf Oz} axes. It is fixed relative to the pulsar companion. The two stars mark the region of the angular emission patterns of highest wave intensity (see section \ref{sec_sources_wind}).
}
\label{radio_AW_cones_emission}
\end{figure}

When the Alfv\'en velocity is close to $c$, as in a pulsar wind,
the total electric current is
\begin{equation} \label{courant_total}
I_{A} \sim 4 (E_0 - E_i) R_b /\mu_0 c
\end{equation}
where $E_i$ is an electric field along the body caused by its ionospheric or surface finite conductivity.
The power dissipated by Joule effect along the body is maximized when internal and external loads match, that is when $E_i=E_0/2$.
For a pulsar with 1 second rotation period and an Earth-like orbiting body at 0.2 UA, $I_{A} \sim 10^{11}$ A. This current 
has the same order of magnitude as the current that powers the whole pulsar magnetosphere \citep{Goldreich_Julian_1969}. For bodies orbiting recycled (millisecond) pulsars, due to a weaker wind magnetic field, the Alfv\'en wing current is smaller than the Goldreich \& Julian current by one or two orders of magnitude ($\sim 10^9$ A), but still not negligible.

With simple models of a pulsar wind, for a pulsar with a magnetic dipole axis aligned with the rotation axis, a radial wind flow and a mostly azimuthal magnetic field in the companion's environment, it is possible to estimate the angle $\delta$ between the invariant vector $\vec{V}_s$, i.e. the direction of the Alfv\'en wing, and the radial direction. 
In most wind models for aligned pulsars (see \citet{Kirk_2009} for a review), the wind is characterized by two invariants along its flow: 
 the neutron star magnetic flux $\Psi$, and the mass flux $f$, defined as
 \begin{equation} \label{eq_f}
f=\gamma_0 \rho'_0 v_0^r r^2 \mbox{ and } \Psi=r^2 B_0^r
\end{equation}
with $\gamma_0$ the Lorentz factor associated to the unperturbed wind velocity $v_0 \sim v_0^r$, and $B_0^r$ the radial magnetic field. Estimates of the Lorentz factor based on various observations and models vary over a fairly large range from $10^{1}$ to $10^{7}$ \citep{Kirk_2009}.
The angle $\delta_s$ of the Alfv\'en wings relative to the local magnetic field is
\begin{equation} \label{eq_delta}
\delta_s=s\arctan{ \left[  \frac{x}{\gamma_0 \left(\sqrt{1+x^2}-s \right)}\right]},
\end{equation}
where $s=\pm1$, and $x$ is a dimensionless parameter defined by
\begin{equation}
x=\frac{r}{\gamma_0 r_{LC}}.
\end{equation}
Details of this computation are given in Appendix \ref{sec_calcul_angle_AW}.

When $x \gg 1$, with Eq. (\ref{eq_gamma_infty}) and 
$\gamma_0 \gg 1$, Eq. (\ref{eq_delta}) reduces to
\begin{equation} \label{eq_delta_infty}
\delta_s \sim s{\gamma_0^{-1}}.
\end{equation}
This means that the two Alfv\'en wings make two symetric and small angles relatively to the radial direction. 
But we have no a priori evidence of the value of $x$. We might as well have $x \ll 1$. In that case, 
the two Alfv\'en wings have an asymetric configuration  with an angle 
$\delta_- = -x / 2 \gamma_0=r_{LC}/(2 \gamma_0^2 r)$ even smaller than in the previous case, and an angle $\delta_+ = \arctan{2/\gamma_0 x}=2 r_{LC}/r$.

A few examples of $\delta_\pm$ angles is displayed in Table \ref{table_x_delta_tau} for various values of $r$ and $\gamma_0$.

\begin{table}
\caption{$x$ ratio, Afv\'en wings angles $\delta_+$ and $\delta_-$ with the radial wind direction, and proper time $\tau_+$ and $\tau_-$ spent by the CMI sources in the wings as functions of the companion distance $r$ and the wind Lorentz factor $\gamma_0$. A few values of $\tau_+$ are in parenthesis because CMI is not expected 
with the $\delta_+$ Alfv\'en wing when $x\ll1$.}
\label{table_x_delta_tau}
\begin{center}
\begin{tabular}{|c| c| c| c| c |c|c|} % centered columns (4 columns)
 \hline % inserts double horizontal lines
 $r$ (AU) & $\gamma_0$    & $x$           & $\delta_+$ ($^\circ$)  & $\delta_-$ ($^\circ$)  &  $\tau_+$/s       & $\tau_-$ (s)       \\ % table heading
 \hline %inserts single line
  0.1      &    10         &    78.        &    5.7           &   -5.6           &   $3.\, 10^{-3}$  &  $3. \, 10^{-3}$  \\
  0.01     &    $10^{3}$   &$7.8\,10^{-2}$ &    1.4           & $-2.\, 10^{-3}$  &   $1. \, 10^{-4}$ &  $8. \, 10^{-2}$  \\ 
  0.1      &    $10^{3}$   &   0.78        &   0.16           &  $-2.\, 10^{-2}$ &   $1.\, 10^{-3}$  &  $1. \, 10^{-2}$  \\
  1.0      &    $10^{3}$   &    7.83       & $6.5\, 10^{-2}$  &  $-5.\, 10^{-2}$ &   $3.\, 10^{-3}$  &  $4. \, 10^{-3}$  \\
 40.       &    $10^{5}$   &    3.13       & $7.8\, 10^{-4}$  &  $-42\, 10^{-4}$ &   $2.\, 10^{-3}$  &  $5. \, 10^{-3}$   \\
 10.       &    $10^{5}$   &   0.78        & $1.7\, 10^{-3}$  &  $-2.\, 10^{-4}$ &   $1.\, 10^{-3}$  &  $1. \, 10^{-2}$   \\
  0.1      &    $10^{5}$   &$7.8\,10^{-3}$ &   0.14           &  $-2.\, 10^{-6}$ &   ($1.\, 10^{-5}$)  &   0.85              \\
  0.01     &    $10^{6}$   &$7.8\,10^{-5}$ &   1.4            &  $-2.\, 10^{-9}$  &  ($1.\, 10^{-7}$)  &  85       \\
 \hline %inserts single line
\end{tabular}
\end{center}
\end{table}

We can consider two kinds of sources of radio-emissions: the source can be (1) attached to the companion body, or more precisely to a plasma with a low velocity in the body's frame of reference; or it can be (2) carried along with the pulsar wind plasma.

\section{Radio emission from a source linked to the companion} \label{sec_sources_attached}

The plasma close to the pulsar companion can be destabilized by the Alfv\'en wing's current flowing along the body, and excite a high level of coherent electromagnetic waves. 
Then, in the reference frame of the companion, the radio emission may be sent at various angles, as it is observed near solar system planets connected to Alfv\'en wings. For instance, Jupiter's galilean satellite Io is embedded in the co-rotating magnetosphere of Jupiter and is the source of two Alfv\'en wings \citep{Neubauer_1980, Saur_2004}. 
 
The wings in the vicinity of Jupiter are strong radio emitters of decametric waves that are observed with ground radio-telescopes. Their frequencies cover the range of local electron cyclotron frequencies in the source regions. 
It is not clear if the sources are fixed relative to the Alfv\'en wings or if they are convected by the co-rotating plasma, but in both cases, their velocity would be low enough for not being measurable by Doppler shift. 

For a pulsar companion, sources of radio waves attached to the companion or to the co-rotating plasma in its vicinity would also have a non-relativistic velocity relative to the observer. The relativistic aberration would be negligible as well, not causing any focussing of the emitted energy. Therefore, even if the source is powerful, these radio waves would be very faint at interstellar distances and the chances to detect them from Earth would be very low. As a comparison, the intense Io-Jupiter radio emission is not detectable against the galactic background beyond 1 pc distance even with the largest existing radiotelescopes \citep{Farrell_2004, Zarka_2004b}. Detection at distances of the order of a kpc would thus require an emission $>10^6$ times more intense.

Because their emission could not be detected from Earth, the assumption of radio sources attached to the companion is not investigated further.

\section{Sources of radio emission carried by the wind along the wings} \label{sec_sources_wind}

Radio emissions could as well be emitted from the plasma wind, when it crosses the Alfv\'en wings, far from the pulsar companion. The sources would be convected with the wind along the Alfv\'en wings. Because the wind is relativistic, the sources would have a relativistic motion relative to an observer on Earth. 
More precisely, we assume that the source propagates along the Alfv\'en wing, and that the source velocity projected along the (radial) wind direction is the same as that of the wind. Then, the source velocity modulus is $V_s = v_0/\cos \delta \sim v_0 (1 + \sin \delta^2)^{1/2}=c^2 (1 - \gamma_0^{-2})(1+\delta^2)$. The requirement that $V_s <c$ implies $\delta^2 < (\gamma_0^2 - 1)^{-1}$. It is satisfied with the two wings if $x\gg1$, and with the $\delta_-$ wing only when $x\ll1$. Thus, at least one source can always exist.

\subsection{Plasma flow, instability, and radio waves} \label{sec_intro_source_mobile}
\begin{figure}
\resizebox{\hsize}{!}{\includegraphics{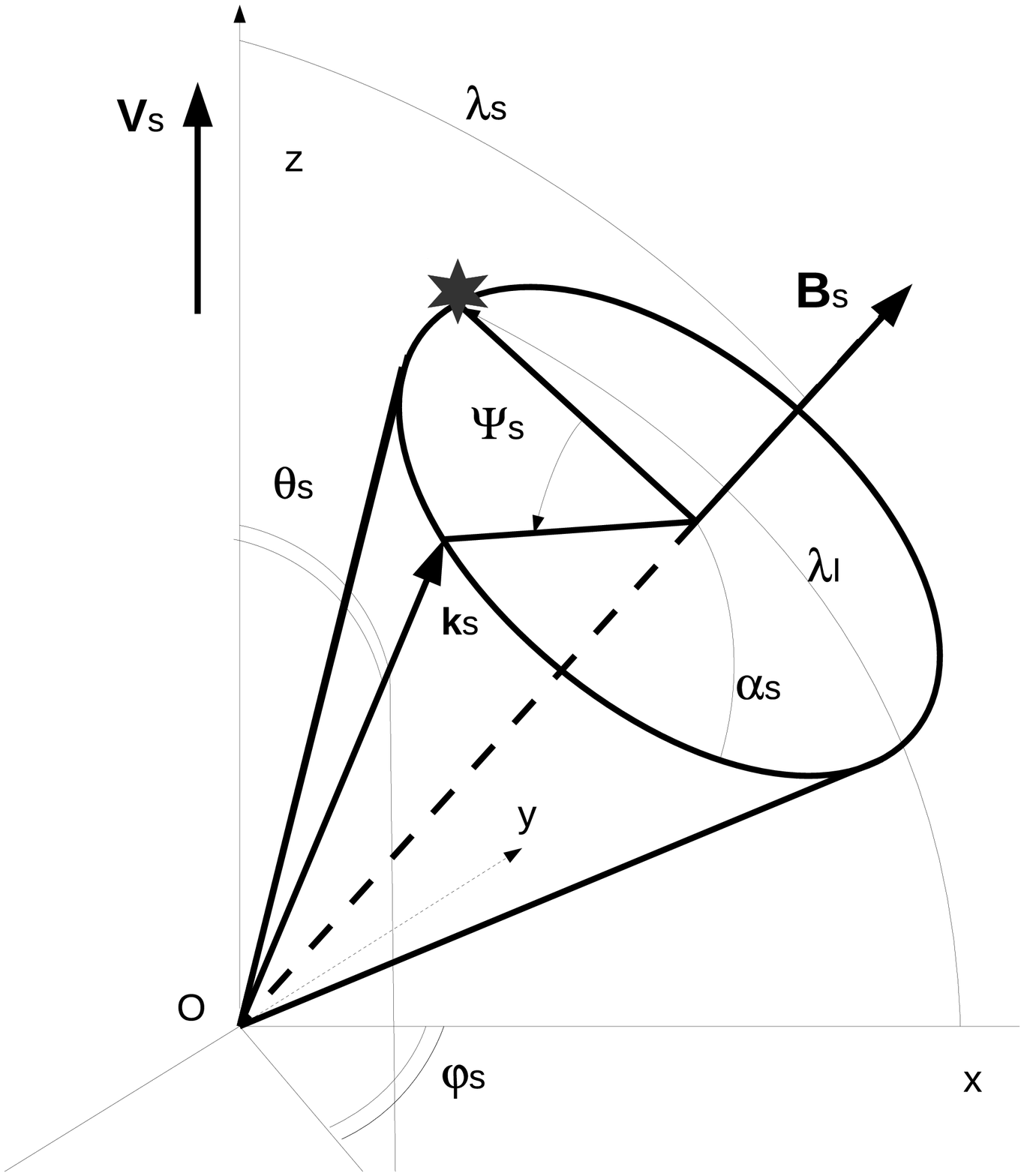}}
\caption{Directions of the magnetic field and of the radio waves emitted in the source frame : the wave vectors are along a cone of summit angle $\alpha_s$, and of axis parallel to the magnetic field $\vec{B}_s$. The wave vector position on that cone is parametrized by the angle $\Psi_s$. The inclination of the magnetic field with the wing axis $Oz$ is $\lambda_s$. The cylindrical angular coordinates $\theta_s$ and $\varphi_s$ of the wave vector are defined relatively to the $z$ axis.
As in Fig. \ref{radio_AW_cones_emission}, the star (at $\Psi_s=0$) marks the angle of emission of the most intense CMI waves.}
\label{shema_principe_diagramme_emission}
\end{figure}

We define the observer's frame by the Cartesian coordinate system where $Oz$ is aligned with the Alfv\'en wing velocity $\vec{V}_s$ and where the magnetic field is contained in the plane $xOz$ (cf. Fig. \ref{radio_AW_cones_emission}). The source frame has the same axes but moves at velocity $\vec{V}_s$ = $V_s \vec{e}_z$ along $Oz$.
The pulsar wind velocity in the observer's frame is $(v_0 \sin \delta, 0, v_0 \cos \delta)$.
The wind velocity in the source frame, noted $\vec{v}'$, is given by
\begin{eqnarray}
v'_z = \frac{v_0 \cos \delta -V_s}{1-\frac{V_s^2 \cos \delta}{c^2}} &\mbox{ and }& v'_x= \gamma v_0 \sin \delta \left[ 1+\frac{V_s}{c^2}v'_z\right]
\end{eqnarray}
where $\gamma = (1-V_s^2/c^2)^{-1/2}= (1-v_0^2 (1+\sin^2 \delta)/c^2)^{-1/2} \sim \gamma_0$.
If the source propagates with the wind, as assumed here, then $V_s = v_0 \cos \delta$, 
$v'_z=0$, $v'_x= v_0 \gamma  \delta= v_0 s x/(\sqrt{x^2+1}-s)$ is a significant fraction of $v_0$.
While in the observer's frame the magnetic field is mostly azimuthal ($B_0^\phi \gg B_0^r$), this is not true in the source frame. In that frame the magnetic field is 
\begin{eqnarray} \label{eq_B_vent_source}
B_s^r = B_0^r \mbox{ and } B_s^\phi = \frac{B_0^\phi}{\gamma} \ll B_0^\phi,
\end{eqnarray}
thus a large component parallel to the wing exists, as illustrated in Fig. \ref{shema_principe_diagramme_emission}.

In terms of particle distribution (a Lorentz invariant), $f(\vec{r},\gamma m \vec{v})=f(\vec{r}',\gamma m \vec{v}')$, so that a distribution that is shifted in the ${v}_z$ direction in the observer's frame is shifted in the $v'_x$ direction in the source frame. The
particle distribution is then highly non gyrotropic, since $Ox$ is a direction making a large angle with the magnetic field is the source frame. This shifted distribution is expected to be a powerful energy source for the cyclotron maser instability (CMI) \citep{Freund_1983,Wu_1985}. 
Let us notice that the CMI theory, even when applied in the middly relativistic plasma of main sequence stars and in the Io Alfv\'en wings, is based on special relativity effects. The CMI does not exist in a non relativistic plasma;  it requires a particle population with non negligible $v/c$.

A computation of the CMI growth rate is not developed in this paper. It would require to know the precise shape of the particle distribution function, for which we have no measurement nor theoretical arguments that would justify any choice at this stage.
Nevertheless, the crucial part of the CMI theory concerns the cyclotron resonance condition. It is developped in the relativistic context without any approximation concerning the Lorentz factor \citep{Wu_1985}. In that general context, the most unstable waves have the following properties in the reference frame of the source.
\begin{itemize}
\item The wave vectors are aligned along a cone of summit angle $\alpha_s$  whose axis is the magnetic field in the source reference frame. 
\item Because the average velocity of the distribution is low in the direction of the magnetic field, the wave vectors are expected to make a large angle with the magnetic field. We can consider that the unstable wave vectors lie along a cone of aperture half-angle $\alpha_{s}$ in the range $85^\circ-90^\circ$.
\item The growth rate does not depend on the azimuth $\Psi_s$ of the wave vector along the cone if the distribution function is gyrotropic (i.e. of the form $f=f(\vec r, p_\parallel, p_\perp)$). On the contrary, the growth rate can depend strongly on $\Psi_s$ if $f$ depends explicitely on the three components of $\vec p$. In the present case, as shown above, the distribution is non gyrotropic, and the intensities of the waves should depend on $\Psi_s$, the most intense waves on this cone corresponding to wave vectors $\vec{k}$ in the $x$ direction, mostly in the direction of the wind ($+x$ for the $\delta+$ wing), and possibly also in the opposite direction.
\item The frequency of the emitted waves is close to the local electron gyrofrequency $\omega \simeq \omega_{ce}$ in the linear approximation ($-x$ for the $\delta+$ wing).
\item When the growth rate is high, non-linear effect tend to produce harmonics and broaden the emitted frequency spectrum.
\end{itemize}

How long does an element of plasma of the wind spend in the Alfv\'en wing ? Let us assume that the diameter of the Alfv\'en wing is $\eta R_b$ where $\eta$ is of the order of one. The inclination angle of the wind relative to the Alfv\'en wing being $\delta = 1/\gamma$, an element of plasma of the wind spends a time $\tau_o \sim \eta R_b / \delta v_0$ in the wing, in the observer's reference frame. In the reference frame of the source, this time is
\begin{equation} \label{eq_duree_passage_dans_source}
\tau_s \sim \eta R_b / \delta \gamma_0 v_0.
\end{equation}
A few values of $\tau_s$ are displayed in Table \ref{table_x_delta_tau} where it is assumed that $\eta R_b=1000$ km, and $v_0 \sim c$.
We can see that these times currently exceed a few milliseconds, especially in the $\delta_-$ wing. 
This is much longer than the typical growth time of the CMI, that can reach saturation in tens to hundreds of microseconds in much less anisotropic and energetic plasmas \citep{Pritchett_1986, Lequeau_1987, Lequeau_1988}.

\subsection{Direction of radio waves emitted by the Cyclotron Maser instability}

Let $\theta_s$ and $\varphi_s$ define the direction of the electromagnetic wave vector in the source frame (Fig. \ref{shema_principe_diagramme_emission}). Because of the relativistic aberration, in the observer's frame, these angles become:
\begin{eqnarray} \label{eq_aberration_angle}
\cos \theta_o = \frac{\cos \theta_s + \beta}{1+\beta \cos \theta_s} \mbox{ and } \varphi_o = \varphi_s
\end{eqnarray}
where $\beta= V_s/c$. 
A well known issue of Eq. (\ref{eq_aberration_angle}) is that for large values of $\gamma$ and for any radiation angle $\theta_s>0$, the corresponding angles $\theta_o$ in the observer's frame are smaller than $\gamma^{-1}$. We will see that this is correct provided that $\gamma \gg 10^3$.
The luminosity of the emission is also affected by the aberration. Let $d W_o/d \Omega_o$ be the energy radiated in the solid angle $\Omega_o$
in the observer's reference frame, and $d W_s/d \Omega_s$ the same ratio in the source frame, then
\begin{equation}
\frac{d W_o}{d \Omega_o} = \frac{d W_s}{d \Omega_s} \frac{(1-\beta^2)^2}{(1-\beta \cos \theta_o)^3} = \frac{d W_s}{d \Omega_s} \gamma^2 (1+\beta \cos \theta_s)^3.
\end{equation}
For an isotropic radiation in the source frame, $\frac{d W_s}{d \Omega_s}$ is constant and the main effect on $\frac{d W_o}{d \Omega_o}$ is the amplification by $\sim \gamma^2$ resulting from the transformation of the solid angle of emission from the source frame to the observer's frame.
For instance, all the radiation for $\theta_s >0$ is radiated in the observer's frame in the solid angle $d \Omega_o \sim \pi \gamma^{-2}$ sr delimited by the cone of summit angle $\gamma^{-1}$, that is very small if $\gamma \gg 1$. Conversely, all the radiation for $\theta_s <0$ is emitted in the other directions 
corresponding to a solid angle $\Omega = \pi (4 - \gamma^{-2})$ sr. 
Therefore, the power angular density $d W_o /d \Omega_o$ is much higher inside the cone of summit angle $\gamma^{-1}$ than in the other directions. 
That is why we can say that most of the observable radiation is contained within the cone of angle $\gamma^{-1}$. 
In the other directions, if $\gamma \gg 1$, the amplitude of the signal is lower by orders of magnitude. 

But theories of the CMI tell us that the radio waves are not emitted isotropically, but rather along conical sheets of aperture angle $\alpha_s$ varying with the frequency in a range close to $90^\circ$. If the particle distribution is non gyrotropic, the intensity of the waves can be highly variable along the cone walls.

We consider first the direction of the wave vectors associated to a single Alfv\'en wing aligned with $Oz$.
Let us consider waves whose vectors $\vec{k}$ form a cone of angle $\alpha_s$ with the magnetic field in the source reference frame. 
How are the wave directions in the observer's reference frame ?
Let $\lambda_s$ be the angle between the emission cone axis and the $z$ direction in the source frame (Fig. \ref{shema_principe_diagramme_emission}). Because the emission cone axis is parallel to the magnetic field, $\tan \lambda_s = r/\gamma r_{LC}$. A single wave vector is characterised by the cone of axis $\vec{B}/B$, of summit angle $\alpha_s$, and by its position along the cone, parametrized by the angle $\Psi_s$. This wave vector direction is characterised by the two angles $\theta_s$ and $\varphi_s$ in the spherical coordinates frame associated to the $Oz$ axis in the source frame. The coordinates of the unit vector parallel to the wave vector are: 
\begin{eqnarray} \label{eq_wave_vector}
 k_x &=& \cos \varphi_s \sin \theta_s = \cos \lambda_I \cos \Psi_s + \\ \nonumber 
 & & \frac{r_n}{1+r_n^2} \left( r_n \cos \lambda_I + \sin \lambda_I \right)(1-\cos \Psi_s) \\ \nonumber
k_y &=& \sin \varphi_s \sin \theta_s = \frac{1}{\sqrt{1+ r_n^2}} \left( \cos \lambda_I - r_n \sin \lambda_I \right) \sin \Psi_s \\ \nonumber
k_z &=& \cos \theta_s = \sin \lambda_I \cos \Psi_s + \\ \nonumber 
 & & \frac{1}{1+r_n^2} \left( r_n \cos \lambda_I + \sin \lambda_I \right)(1-\cos \Psi_s)
\end{eqnarray}
where $\lambda_I=\pi/2 -\lambda_s + \alpha_s$ and $r_n=r/\gamma r_{LC}=\tan \lambda_s$.
The direction angles $\theta_s$ and $\varphi_s$ can be easily derived and expressed in the observer's frame following Eq. (\ref{eq_aberration_angle}).
When $\beta$ is close to one, it is better to use the following relation that resolves a problem of numerical degeneracy,
\begin{equation} \label{eq_aberration_angle2}
\theta_o = \frac{1}{\gamma}\sqrt{\frac{1-\cos \theta_s}{1+\beta \cos \theta_s}}.
\end{equation}
The above angles are relative to a single Alfv\'en wing, and the $z$ axis corresponds to the wing direction. 

As mentioned, the luminosity is expected to be much higher in the direction corresponding to $\Psi_s=0$.

\begin{figure}
\resizebox{\hsize}{!}{\includegraphics{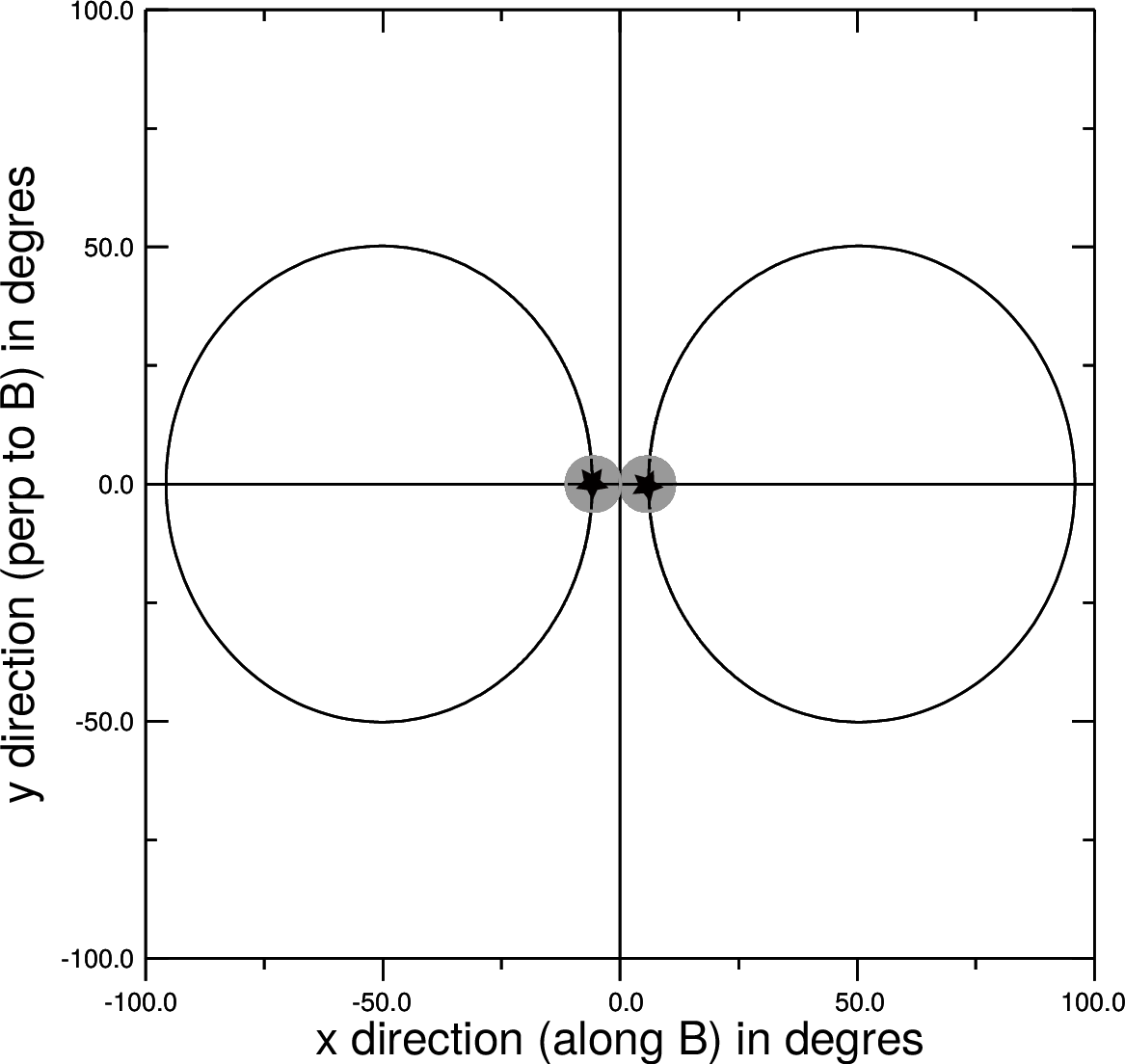}}
\caption{Directions of the radio waves in the observer's frame.  The two smaller tangent grey disks %of radius $\delta$ 
in the middle of the plot correspond to the waves that are beamed at $\leq 90^\circ$ from the  Alfv\'en wing axis in the source frame in the approximation of very large $\gamma$ values (see text for the precise meaning of "very large"). 
%The external circle limiting each grey disk corresponds to emission at $90^\circ$ from the  Alfv\'en wing axis in the source frame. 
The two large circles (black lines) are the radio beaming patterns associated to the CMI. They are defined by $x=  \delta_\pm + \theta_o \cos \varphi_o$ and $y=\theta_o \sin \varphi_o$, with $\theta_o$ and $\varphi_o$ the cylindrical angular coordinates of the wave vector in the observer's reference frame of each Alfv\'en wing. $\theta_o$ and $\varphi_o$ are related via Eq. (\ref{eq_aberration_angle}) (or Eq. (\ref{eq_aberration_angle2})) to the wave vector coordinates in the source frame $\theta_s$ and $\varphi_s$ given by Eq. (\ref{eq_wave_vector}). The most intense emissions correspond to the innermost intersection of each large circle with the horizontal axis. Here $\gamma=10$, $r = 0.1$ AU and $\alpha_s=85^\circ$. With these parameters, $x=78$. The stars mark the directions of the most intense radio emissions.}
\label{fig_diagramme_rayonnement_gamma_10}
\end{figure}

When the two wings are considered, it is better to define a $Z$ axis along the wind direction, and to shift the above values of $\theta_o$ (related to a wing $z$ axis) by an angle $\delta_-$ for one wing and $\delta_+$ for the other. The directions of the waves are plotted in Fig. \ref{fig_diagramme_rayonnement_gamma_10} in the observer's frame for a planet at $0.1$ AU from the star, an emission angle $\alpha_s=85^\circ$, and $\gamma=10$ (the low range of $\gamma$ for a pulsar wind). The small grey disks represent the field illuminated by waves emitted at an angle $\leq 90^\circ$ from each  Alfv\'en wing axis in the source frame (cf. Fig. \ref{shema_principe_diagramme_emission}) in the approximation of very large $\gamma$. 
There is one such disk for each Alfv\'en wing, whose border corresponds to emission at $90^\circ$ from the  Alfv\'en wing axis in the source frame. 
These circles are the same that limit the cones represented on Fig. \ref{radio_AW_cones_emission}. %and \ref{emission_AW_vue_lobes}. 
Because these grey disks do not encompass the other circles, we see that the approximation of a very large $\gamma$ is not valid when $\gamma =10$. 

The angular emission patterns can be related to time intervals. Let us first neglect the oscillation of the (tilted) magnetic field with the pulsar spin frequency. 
The simplest trajectory of the line of sight to the observer on these figures is a straight line uniformly parametrized in time, deduced from the orbital motion of the pulsar companion. In Fig. \ref{fig_diagramme_rayonnement_gamma_10}, there would be four intersections of the emission beam (large circles) with this straight line, two of which being associated to high amplitude signal. An angle interval $\Delta$ corresponds to a time interval $\tau$ given by
\begin{equation} \label{eq_largeur_pulse_gamma_Torb}
\left(\frac{\mbox{deg}}{\Delta}\right) \left(\frac{\tau}{\mbox{s}}\right)=240 \,\left(\frac{T_{orb}}{ \mbox{day}}\right).
\end{equation}
When the oscillation of the magnetic field is taken into account, the whole figure rotates back and forth around the origin at the pulsar period with an angular amplitude depending mainly on the pulsar tilt angle. Then, provided that the time $\tau$ is larger than the pulsar spin period, we can have several pulses for each crossing of the beam pattern, instead of a longer one when the oscillation is negligible.

For $\gamma=10$, the angle $\lambda_s$ between the cone axis and $Oz$ in the source frame is not small, so that a large part of the emission cone of summit angle $85^\circ$ is beamed at $>90^\circ$ from $Oz$. In the observer's frame, we see on Fig. \ref{fig_diagramme_rayonnement_gamma_10} that these radio emissions make a large angle -- of tens of degrees -- with the wind direction. Due to the orbital motion of the pulsar companion, a distant observer will cut the emission diagram along a horizontal line shifted from the horizontal axis of Fig. \ref{fig_diagramme_rayonnement_gamma_10} by the inclination of the companion's orbital plane as seen from Earth. If the pulsar magnetic field is tilted on its rotation axis, then the beaming pattern will oscillate at the pulsar spin period (and thus the observer will cut the emission diagram along a sinusoidal curve). In all cases, the observer has a good chance to intersect four times the radio emission diagram during a full companion's orbit. But, because of the relativistic transformation of solid angles and because $\Psi_s = 0$ corresponds to much more powerful emissions than in the other directions, the most intense emissions will be seen at the crossings of the radio beaming diagram close to the origin, inside the grey disks.

\begin{figure}
\resizebox{\hsize}{!}{\includegraphics{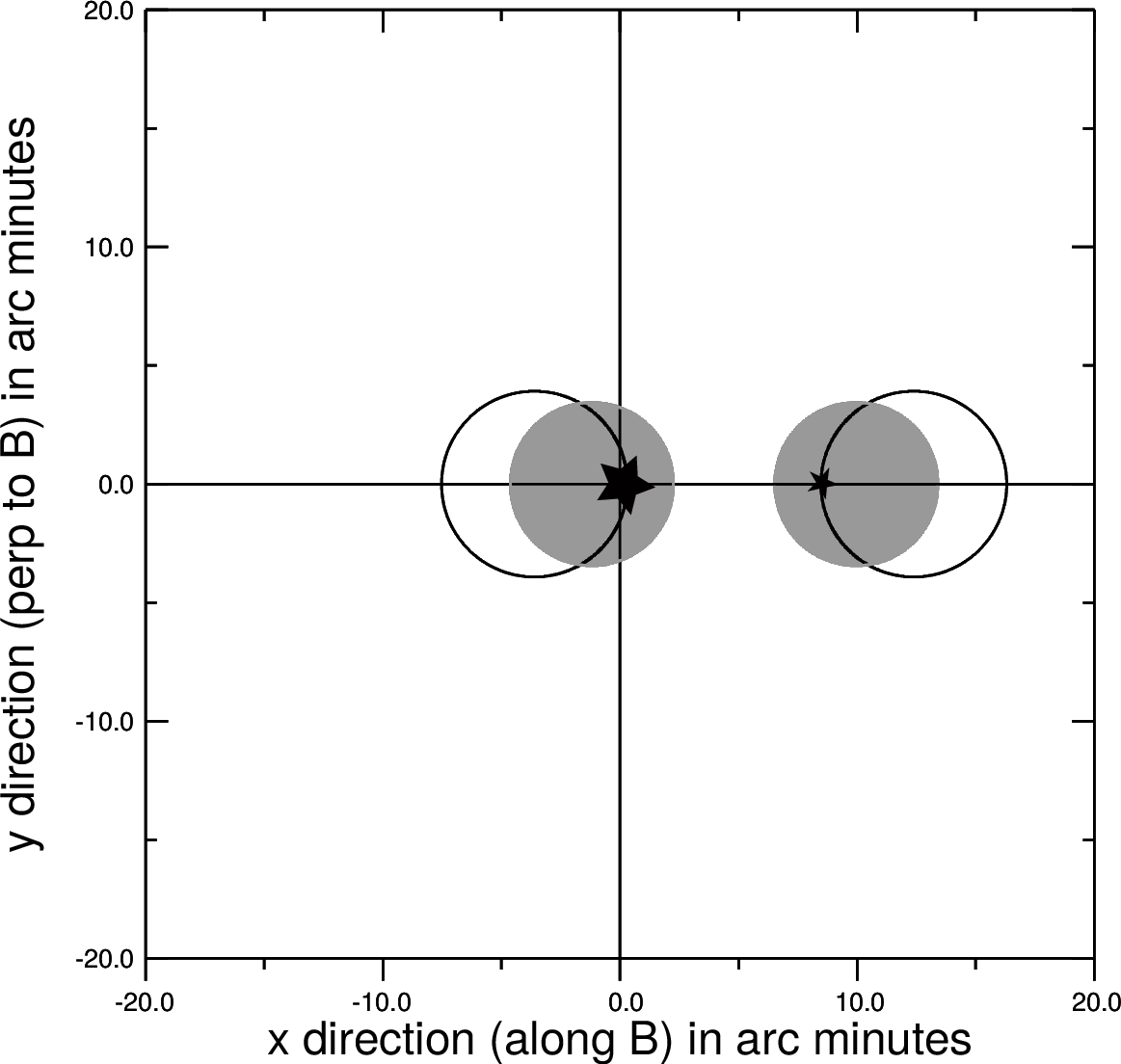}}
\caption{Directions of the radio waves in the observer's frame. Same as in Fig. \ref{fig_diagramme_rayonnement_gamma_10} but for $\gamma=1000$. As in Fig. \ref{fig_diagramme_rayonnement_gamma_10}, $r = 0.1$ AU and $\alpha_s=85^\circ$. These parameters correspond to $x=0.78$.}
\label{fig_diagramme_rayonnement_gamma_1000}
\end{figure}

The situation is quite different for $\gamma=1000$ as shown in Fig. \ref{fig_diagramme_rayonnement_gamma_1000}. The range of angles of emission is now much less than $1^\circ$. Therefore, the chances of being in the right line of sight are much lower. But if it is so, the observer can see the radio emission up to four times (out of which two correspond to intense emission) that are much closer in time to each other. At a distance of $0.1$ AU from a pulsar with a mass of 1.4 $M_\odot$, the orbital period is $\sim$ 10 days, and the typical separation of 0.16$^\circ$ between the two innermost intersections of the radio beaming pattern correspond to time intervals of $\sim$6 minutes. Because of the already mentioned anisotropy of the distribution function, the two inner regions (near the two inner intersections with the horizontal line) correspond to much higher amplitudes than the two others.

\begin{figure}
\resizebox{\hsize}{!}{\includegraphics{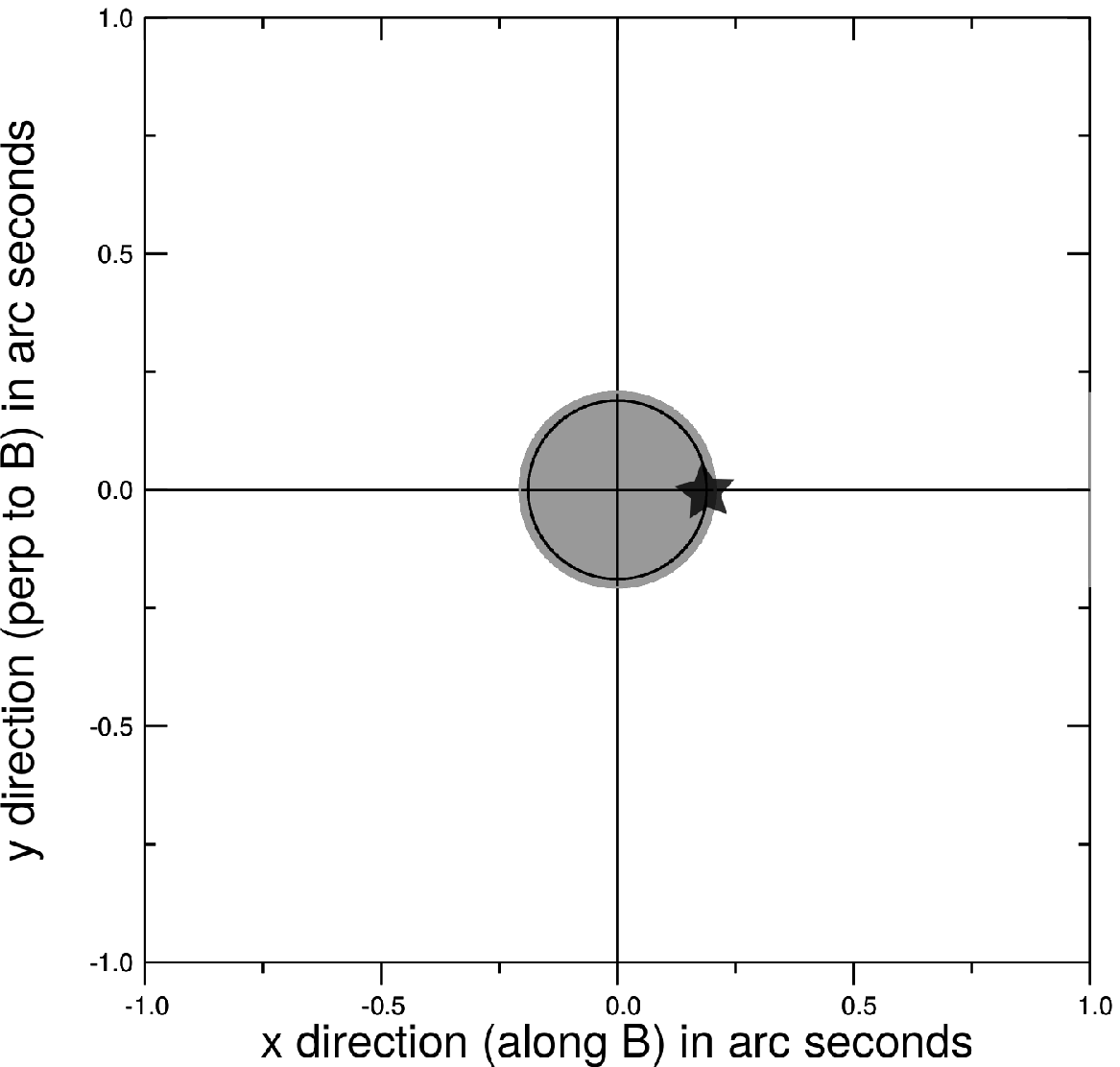}}
\caption{Directions of the radio waves in the observer's frame. Same as in Fig. \ref{fig_diagramme_rayonnement_gamma_10} but for $\gamma=10^6$ and $r = 0.01$ AU. As in Fig. \ref{fig_diagramme_rayonnement_gamma_10}, $\alpha_s=85^\circ$. These parameters correspond to $x=0.78 \, 10^{-5}$. Here only the emission pattern associated to $\delta_-$ is represented. The star marks the direction of the most intense radio emissions. }
\label{fig_diagramme_rayonnement_gamma_1e6_0p01UA}
\end{figure}

For $\gamma=10^6$, only the emission pattern associated to $\delta_-$ is shown on Fig. \ref{fig_diagramme_rayonnement_gamma_1e6_0p01UA}. 
As we can see in Table \ref{table_x_delta_tau}, the time spent by the CMI sources in the wing $\delta_-$ is 85 s, while it is about 0.1 $\mu$s in the other wing.
Furthermore we have seen that a CMI source cannot propagate at the wind velocity along the $\delta_+$ wing when $x\ll1$.   The directions of emission associated to the $\delta_-$ wing almost coincide with the  tangent circle limiting the grey disk (they coincide exactly for $\alpha_s=90^\circ$) and the corresponding angles are now of the order of $0.2''$. We have chosen a distance $r=0.01$ AU, corresponding to $T_{orb} \sim 9$h (with a neutron star of 1.4 solar mass). According to Eq. (\ref{eq_largeur_pulse_gamma_Torb}) the time interval between two circle crossings is less than 5 ms. The vicinity of the intersection of the circle with the horizontal line, on the right-hand side, corresponds to the highest wave amplitude.

For $\gamma=10^5$ and $r=0.1$ UA, the figure (not shown) has the same aspect as in Fig. \ref{fig_diagramme_rayonnement_gamma_1e6_0p01UA}, with an anglular diameter $\sim 2''$  instead of $0.2''$. This corresponds to a time interval of 1.3 second.

Because most pulsar wind models are based on highly relativistic flows, and because, as we will see below, it may explain some already existing observations, we focus the forthcoming analysis on the cases where $\gamma > 10^3$. 

\subsection{Range of frequencies} \label{section_frequencies}

As mentioned in section \ref{sec_intro_source_mobile}, we first consider that the waves are emitted at frequencies close to the local gyrofrequency $f_{c,s}$ in the reference frame of the sources. Then we convert the frequency to $f_{c,o}$ into the frame of the observer.
Using Eqs. (\ref{eq_b0phi}) and (\ref{eq_B_vent_source}) and considering that $\Psi=B_* R_*^2$, the gyrofrequency in the source frame is 
\begin{equation}
f_{c,s}=\frac{q B_*}{2 \pi m}\frac{ R_*^2}{r^2} \sqrt{1+\left( \frac{r}{\gamma r_{LC}} \right)^2}
\end{equation}
with $q$ and $m$ the electron's charge and mass.
The Doppler-Fizeau shift along the line of sight towards the observer provides the observed frequency 
\begin{eqnarray} \nonumber
f_{c,o}=25 &\gamma& \left(\frac{B_{*}}{10^5 \mbox{T}} \right)
 \left(\frac{1 \mbox{AU}}{r}\right)^{2}
 \left( \frac{R_*}{10^4 \mbox{m}} \right)^{2}
 \times 
 \\
& & \left\{1+ \left[\frac{\pi \; 10^5}{\gamma} \left(\frac{10 \mbox{ms}}{T_{*}}\right)\left(\frac{r}{1 \mbox{AU}}\right) \right]^2\right\}^{1/2}.
\label{eq_omega_observer}
\end{eqnarray}

\begin{figure}
\resizebox{\hsize}{!}{\includegraphics{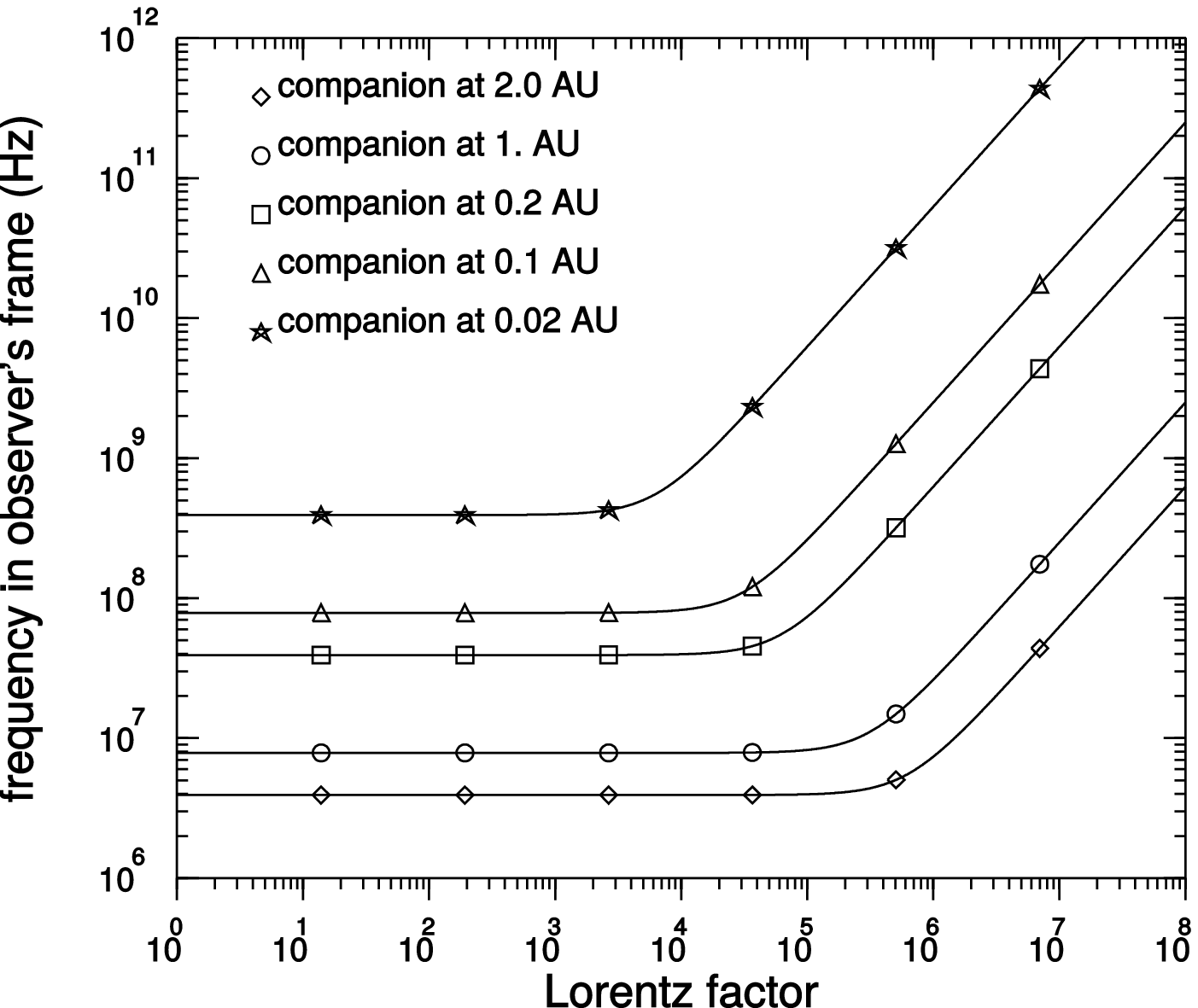}}
\caption{The cyclotron frequency in the observer's frame as a function of the pulsar wind Lorentz factor $\gamma$, from Eq. (\ref{eq_omega_observer}) for various distances of the companion. Values that are not varied are the same as in the text: the magnetic field is $B_*=10^5$ T and the spin period $T_*=10$ ms.}
\label{fig_omega_vs_gamma}
\end{figure}

Figure \ref{fig_omega_vs_gamma} displays the observed frequency $f_{c,o}$ as a function of the Lorentz factor $\gamma$ of the pulsar wind for various companion distances from a millisecond pulsar. 
For $\gamma < 10^5$ the frequencies are in the range of meter to decameter radio waves, typically reachable with LOFAR \citep{vanHaarlem_2013} and (in the future) SKA (www.skatelescope.org). 
The GHz range of frequencies is reached for companions within 0.2 AU when the Lorentz factor is higher than $10^{6}$. 
The sub-millimeter range is reached only for very close planets and very fast pulsar winds ($\gamma>10^7$). 

\begin{figure}
\resizebox{\hsize}{!}{\includegraphics{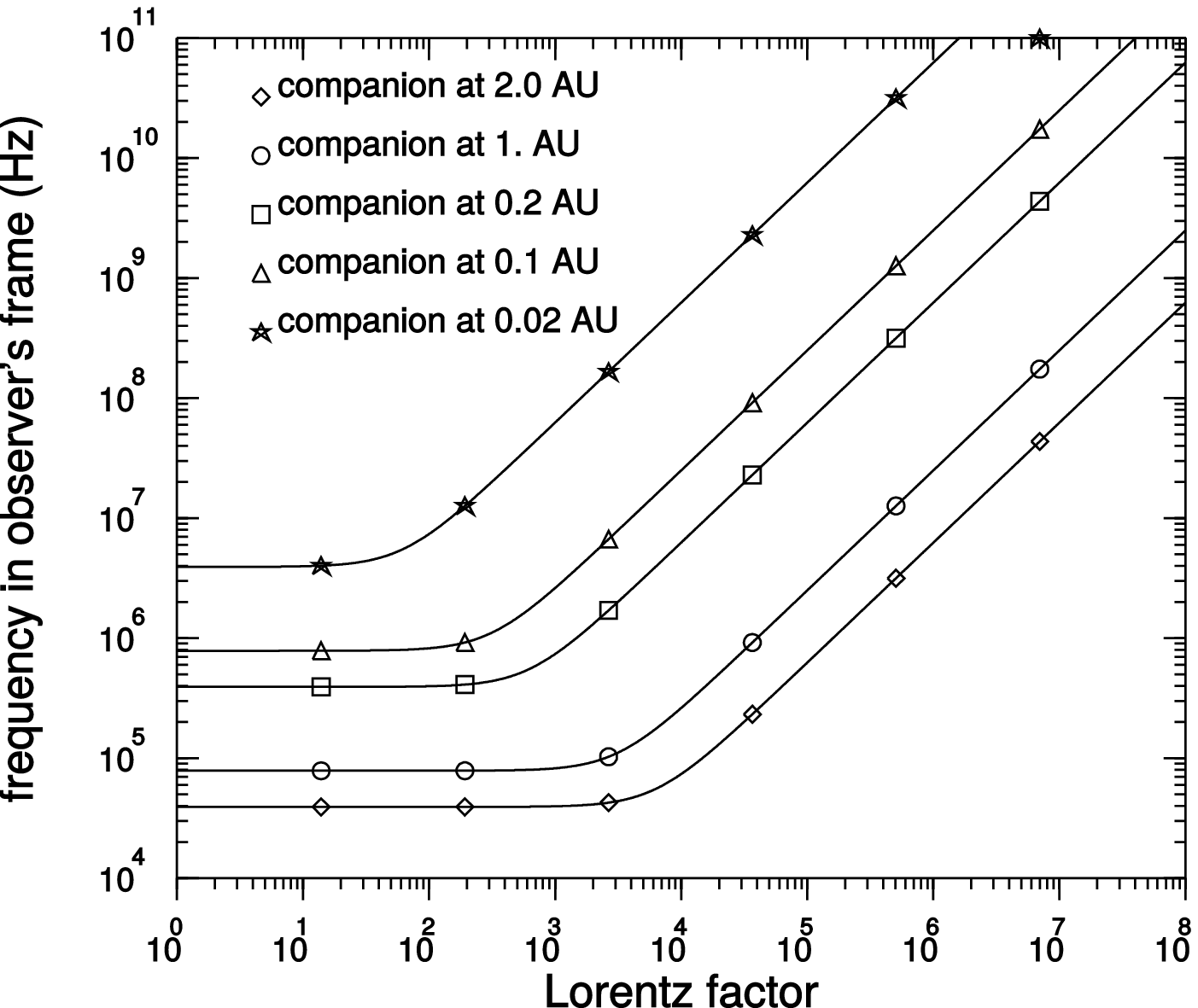}}
\caption{Same as Fig. \ref{fig_omega_vs_gamma} for companions orbiting a standard pulsar with a magnetic field $B_*=10^8$ T and a spin period $T_*=1$ s.}
\label{fig_omega_vs_gamma_PSR_1s}
\end{figure}

Figure \ref{fig_omega_vs_gamma_PSR_1s} displays the observed frequency/Lorentz factor relation for a standard pulsar of period $T_*=1$s and magnetic field $B_*=10^8$ T. Up to now, no planet is known to orbit a standard pulsar, but other companions are likely to exist. For such systems, observed frequencies will be significantly higher than those associated to millisecond pulsar companions. They will remain in the radio range for distant companions, but could reach the infrared domain for close-by companions and high speed winds. 

\subsection{Brightness} \label{section_brightness}

The maximum power involved in an Alfv\'en wing is \citep{Mottez_2011_AWW}
\begin{equation}
\dot E_J = \frac{\pi c}{\mu_0} R_b^2 (B_0^\phi)^2.
\end{equation}
This expression is the relativistic analogue of that derived in \citep{Zarka_2001} for magnetized flow-obstacle interactions.
From Eq. (\ref{eq_b0phi}), this can be expressed as a function of the distance $r$ and the radius $R_b$ of the orbiting body
\begin{equation}
\dot E_J = \frac{\pi}{\mu_0 c} {R_b}^2{r}^{-2} R_*^4 B_*^2 \Omega_*^2.
\end{equation}
From studies of planetary magnetospheric radio emissions in our solar system, it was found that the emitted radio power is proportional to the Poynting flux of the
Solar wind on the magnetospheric cross-section. The same relation of proportionality also relates satellite-induced (Io, Ganymede) radio emissions in the magnetosphere of Jupiter with the incident Poynting flux convected by the rotation of the planetary magnetic field on the obstacle (ionosphere of Io, magnetosphere of Ganymede). The
general scaling law describing this relation has a constant of proportionality $\epsilon \sim 2-10 \times 10^{-3}$ \citep{Zarka_2001, Zarka_2007}. 

In order to test the possible saturation of this scaling law toward large Poynting fluxes and high radio powers \citet{Zarka_2010} analyzed the literature on radio emission from magnetized binary stars. \citet{Budding_1998, Richards_2003} showed that the RS CVn system V711$\tau$ frequently emits radio bursts reaching 0.1 Jy at 2.3 GHz and 1 Jy at 8.4 GHz. With a distance of 29 pc and a bandwidth $\Delta f \geq 8$ GHz, one deduces an isotropic radio power $\sim 7 \times 10^{19-20}$ W. Based on estimates from \citet{Saar_1996, Budding_1998}, RS CVn magnetic fields reach $1-3\times10^3$ G, which leads, based on the separation of the system V711$\tau$ (11.5 $R_\odot$ \citep{Richards_2003}), to a magnetic field magnitude $B \sim 10-30$ G at the interaction region. The radius of this region is at most that of the secondary companion in V711$\tau$ ($\sim 1.3$ $R_\odot$). Assuming a stellar wind velocity $V \sim$300 km/s and an interaction region size $R \sim 1-9 \times 10^8$ m, the Poyinting flux convected on the obstacle is
$\frac{VB^2}{\mu_\circ} \pi R^2 = 7\times 10^{21} - 6\times 10^{24}$ W. The constant of proportionality $\epsilon$ is thus in the range $10^{-1}$ to $10^{-5}$, but more reasonably (matching high and low estimates) in the range $10^{-2}$ to $10^{-4}$. In spite of the above uncertainties, the range of $\epsilon$ falls remarkably close to that deduced from the above scaling law, suggesting that it may remain valid with a constant $\sim$ a few $10^{-3}$ up to 10 orders of magnitude above
the solar system planets range.

Therefore, we estimate that the power radiated in the form of non-thermal radio waves is in our case
\begin{equation} \label{eq_radio_power}
P_{radio}= \epsilon \dot E_J = \epsilon \frac{\pi}{\mu_0 c} {R_b}^2{r}^{-2} R_*^4 B_*^2 \Omega_*^2
\end{equation}
with a conservative value of the order of $10^{-3}$ for $\epsilon$.

At a distance $D$ from the source, for a spectrum of emitted waves spreading along an interval $\Delta f$ of frequencies, the average flux density in Jy of radio waves inside the cone of emission is
%$<S>=4 \gamma^{2} P_{radio} D^{-2} \Delta f^{-1}$. Combining the above equations, the flux density in Jy is
\begin{eqnarray} \nonumber
\left(\frac{<S>}{\mbox{Jy}}\right) &=& 6.5 \left( \frac{\gamma}{10^5}\right)^2 \left(\frac{\epsilon}{10^{-3}}\right) \left( \frac{R_b}{10^7 \mbox{m}} \right)^2 \left( \frac{1 \mbox{AU}}{r} \right)^{2} \left( \frac{R_*}{10^4 \mbox{m}} \right)^4 \times 
\\ \label{eq_flux_density_reduced}
 & & \left( \frac{B_*}{10^5 \mbox{T}} \right)^2 \left(\frac{10 \mbox{ms}}{T_{*}}\right)^2 \left( \frac{Mpc}{D} \right)^2 \left(\frac{\mbox{1 GHz}}{\Delta f}\right).
\end{eqnarray}
The detected flux density will be $S=<S>$ only if the radiation is isotropic in the source frame. If this is not the case, an additional amplification factor intrinsic to the emission mechanism is to be applied for specific directions of emissions inside the cone, whereas $S$ can be null outside of these specific directions. 

\begin{figure}
\resizebox{\hsize}{!}{\includegraphics{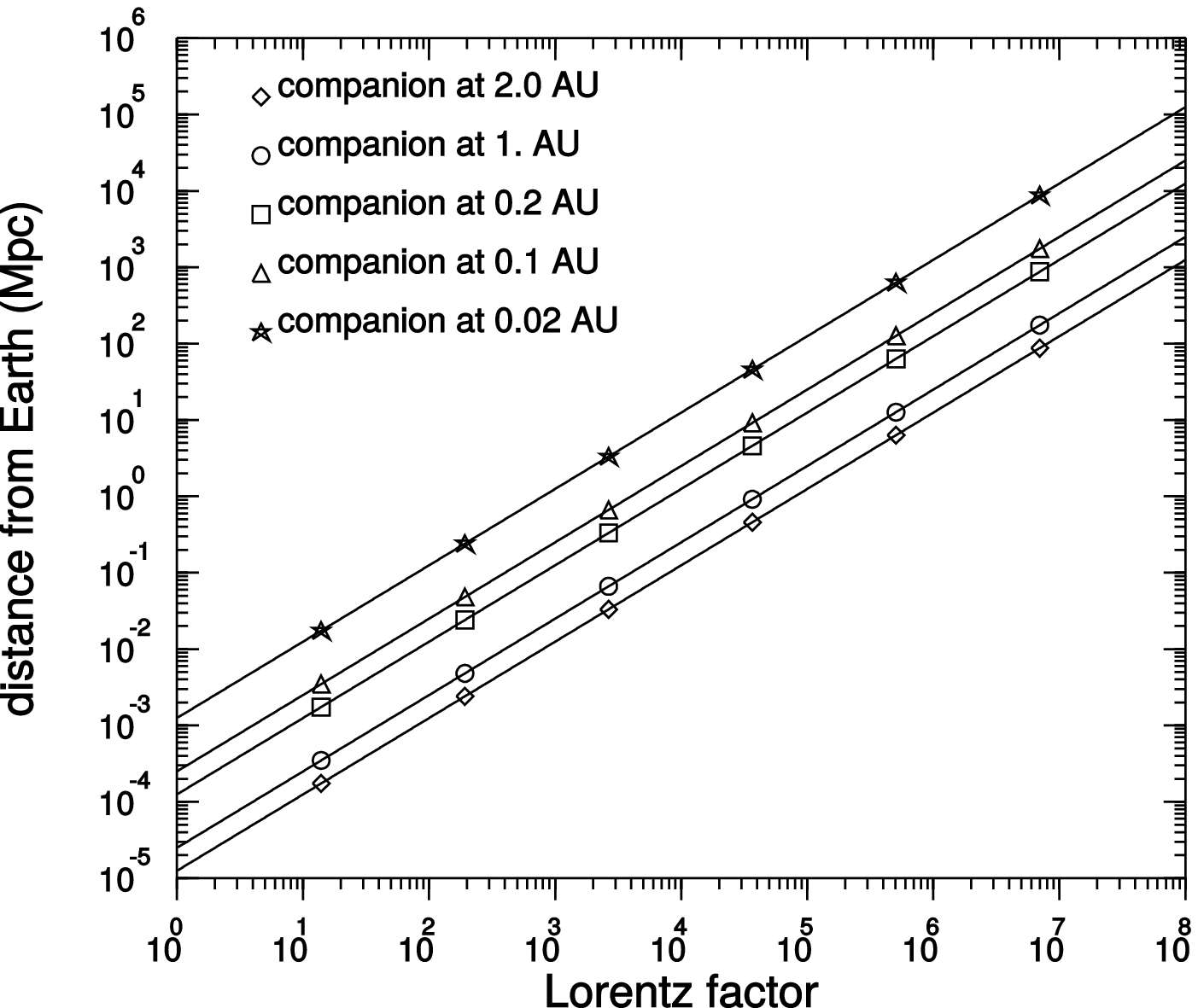}}
\caption{The distance, expressed in Mpc, at which the radio emission produced in the companion's Alfv\'en wing can reach a flux density of 1 Jy, for various distances from the pulsar to its companion. The values that are not varied are the same as in the text. They correspond to a recycled millisecond pulsar, and a coefficient $\epsilon=10^{-3}$. The radio spectrum is assumed to be spread over a spectral range of 1 GHz.}
\label{fig_D_vs_gamma_MS_PSR}
\end{figure}

Figure \ref{fig_D_vs_gamma_MS_PSR} shows, for the reference values of Eq. (\ref{eq_flux_density_reduced}),
the distance $D$ at which the radiation can be observed as a function of the Lorentz factor $\gamma$ of the wind, for various distances between the neutron star and its companion. We can see that for Lorentz factors larger than $\sim 10^6$, the radio emission can be observed as far 1 Gpc or more. For a radio emission produced by e.g. the CMI, an additional amplification factor $\geq 10$ (already in the source frame) is to be taken into account relative to the isotropic power derived above (see \citet{Zarka_2004c} and references therein). Then, a Lorentz factor of $\sim 10^5$ allows detection of 1 Jy emission at 1 Gpc range. Within reasonable ranges for the parameters in Eq. (\ref{eq_flux_density_reduced}), it is possible to produce 1 Jy radio emission at $>$1 Gpc range even for $\epsilon=10^{-4}$ and not excessively high $\gamma$ factors.

The power $P_{radio}$ radiated in the form of radio waves in the companion's wake %, estimated in Eq. (\ref{eq_radio_power_reduced}) 
of the order of $10^{18}$ W for the nominal values of the computation ($10^{20-22}$ W for a companion at 0.1$-$0.01 AU), is weaker than or comparable to the radio power typically emitted by isolated pulsars ($\sim 10^{20}$ to $10^{22}$ W as deduced from the ATNF Pulsar Catalogue \citep{Manchester_2005a, Manchester_2005b}). But the flux density is considerably higher for the companion-induced emission, due to the very high degree of collimation of the radio beams, caused by the highly relativistic motion of the sources, and leading to an amplification of the flux density by a factor $\sim \gamma^2$. It is this high collimation, and not the intrinsically high power of the source, that permits their detection at very large distances. 

Somehow, this phenomenon presents analogies with the high intensity of BL Lac galaxies : they are very bright because we (observers) are aligned with a collimated beam, and because of the relativistic motion of the sources. In the case of BL Lac, the local emission is caused by incoherent synchrotron  and inverse Compton radiation at frequencies largely above the cyclotron frequency. In the present case, we expect coherent plasma process (like the CMI) generating a radiation that, in the frame of the source, is at the local cyclotron frequency. We have not considered the possibility of a high energy counterpart associated to synchrotron radiation from a hot plasma component, but this counterpart, emitted by an incoherent process, would be much less powerful. (Jupiter's synchrotron emission if five orders of magnitude weaker than its auroral CMI radio emission \citep{Zarka_2004a}.)

\section{Possible explanation of remarkable radio transients} \label{possible_explanation}

\subsection{The Lorimer burst and fast radio pulses}

We propose that the Lorimer burst and FRBs described in the introduction may be the signature of the above-described radio emission from pulsar companions. The inferred and surprisingly large distances deduced in \citet{Lorimer_2007} (0.5 Gpc), and even more in \citet{Thornton_2013} (1.7 -- 3.2 Gpc) are consistent with the energetics of the pulsar companion signal discussed in the previous section. The corresponding sources would indeed lie in other galaxies, at Gpc distances from us, but as explained above the exceptionally high collimation of the radio emission allows us to detect these bursts with flux densities of several Jy. This high collimation is also consistent with the rare detection of FRBs, although we do not address this point quantitatively here (see rate calculations in \citep{Hassall_2013}). We discuss here the duration and frequency range of these bursts.

Let us first consider the duration of FRB signals, that we will take as $\sim$ 5 ms (part of which may be due to scatter--broadening (\citep{Thornton_2013}). 
From Fig. \ref{fig_D_vs_gamma_MS_PSR}, it appears that the Lorentz factor $\gamma_0=10^6$ would allow for a detection at Gpc distance. The example displayed in Fig. \ref{fig_diagramme_rayonnement_gamma_1e6_0p01UA} shows that a companion with an orbital period of a few hours, in a $\gamma_0=10^6$ wind could provide a signal of adequate duration.

A value of $T_{orb}$ of order of or larger than a few hours is a reasonable value for the orbital period of a pulsar companion:   more than 70 pulsar companions are known with $T_{orb}<1$ day, 10\% of which have $T_{orb}<0.1$ day \citep{Manchester_2005a, Manchester_2005b}), not including possible low-mass planets and asteroids. Moreover, the value of $T_{orb}$ deduced above is consistent with the fact that no other pulse was observed during the 1.5 hours following the observation of the Lorimer burst and other FRBs. 

Considering the asymetry of the intensity diagram along the circular emision pattern of Fig. \ref{fig_diagramme_rayonnement_gamma_1e6_0p01UA}, the crossing of a pattern corresponding to $\gamma \sim 10^5$ is also compatible with the observations, provided that the peak signal from one wing largely dominates whereas the other is below the detection threshold.

If the emission is produced by a coherent plasma process such as the CMI, we expect that the radiation is produced along the Alfv\'en wings at the local cyclotron frequency.  
Figures \ref{fig_omega_vs_gamma} and \ref{fig_omega_vs_gamma_PSR_1s} show that we can easily explain an emission at $\sim$1 GHz. But the large spectral range over which FRBs are observed ($\geq 25$\%) implies that the emission should be produced along the Alfv\'en wings across a large radial range over which the wind magnetic field and/or $\gamma$ factor vary in this proportion. Another, perhaps more appealing, explanation is that the shift of the velocity distribution of the whole wind along the $v'_x$ direction in the source frame (i.e. the perpendicular direction, at a particular azimuth) provides an enormous free energy to the CMI, far beyond what is usually available in planetary magnetospheres where a small fraction of the particles are gyrotropically distributed at $v_{\perp} \ne 0$. As a consequence, the emission excited by a pulsar companion could be extremely saturated, generating series of harmonics possibly resulting in a quasi-continous broad spectrum.
Actually, one of the FRBs reported in \citep{Thornton_2013} indeed shows 100-MHz-wide bright bands.

\subsection{An event in the Galaxy} \label{sec_PSR1928p15}

\citet{Deneva_2009} reported the discovery of an event composed of three pulsar like bursts from a source named PSR J1928+15. The bursts occured at intervals of 0.403 s from each other, with the middle pulse being brighter by an order of magnitude than the other two. The duration of each of the three dedispersed pulses is of the order of 30 ms. The dispersion measure is DM=242 ps cm$^{-3}$ which, for the galactic coordinates of PSR J1928+15 (longitude=50.64$^\circ$, latitude=$-1.03^\circ$), is compatible with a signal from our Galaxy. In spite of several follow-up observations, the source was not detected again. \citet{Deneva_2009} suggest -- after \citet{Cordes_2008} -- that it could be caused by the excitation of a dormant pulsar by accretion of material from an asteroid belt. But they don't provide any physical insight on the underlying mechanism, while \citet{Cordes_2008} were mostly concerned with pulsar nullings, rather than with isolated pulses from dormant pulsars.

Let us consider a companion orbiting a pulsar with a $P_*=0.403$ s period, in a $\gamma_0=10^3$ wind at a distance $r=0.01$ AU. From Table \ref{table_x_delta_tau},
we can see that (for a $\sim 1000$ km companion), the CMI sources spend $\tau_-=80$ ms in the $\delta_-$ wing, and only $\tau_+=0.1$ ms in the other. Therefore, the CMI develops only in the $\delta_-$ wing. Because $\tau_-/\tau_+$ does not depend on the size $R_b$ of the body, this is true for a smaller object as well.  The high amplitude part of the angular pattern caused by the particle distribution anisotropy of the $\delta_-$ Alfv\'en wing could be crossed in about 1.5 seconds, and this would correspond to 3 pulses of 0.403 seconds. Each of these pulses would correspond to a slightly different value of $\alpha_s$, the second pulse (centre of the emission pattern) being the most intense. Other weaker pulses might have been seen with a lower noise level. The low amplitude part of the emission angular pattern of the $\delta_-$ wing would not have been seen either. Of course, the parameters $\gamma_0=10^3$ and $r=0.01$ AU are simply one possibility among many others. With a single observation, it is not possible to constrain the model more strictly. Nevetheless, we can see that the PSR J1928+15 signal is compatible with our pulsar companion model.

PSR J1928+15 is thought to be at a distance $\sim$10 kpc, much closer than the sources of the Lorimer burst and FRBs. However, the signal amplitude ($S_{peak} \sim$ 180 mJy) is slightly weaker than that of the very distant FRBs. Therefore, the radio luminosity of this system is much lower. A possible explanation is that its orbiting companion is much smaller than those associated to the extragalactic FRBs. For instance, an asteroid (of radius $\sim$ km) would be a possible companion of such a standard pulsar. Because of the companion smaller size and according to Eq. (\ref{eq_duree_passage_dans_source}) the wind plasma would spend less time in the Alfv\'en wings, allowing less time for the full development of the instability that cause the radio emissions. Because the initial phase of the instability growth is exponential, a variation of this duration by a factor of a few units can imply orders of magnitude difference on the luminosity. The radio signature of such a small companion would not be detectable at extragalactic distances.

\section{Discussion and conclusion} \label{sec_conclusion}

\subsection{Why we do not necessarily see the radio emission from the central pulsar} \label{sec_not_seen}

According to our model, the radio emissions from pulsar companions are beamed along the pulsar--companion line. Because the orbital plane of the companion is likely to be very close to the pulsar's equatorial plane, the radio emission from the companion is beamed nearly at right angle from the rotation axis of the neutron star. Because the radio signals from a pulsar are emitted from its polar cap, a very large tilt of the magnetic field is needed to be able to see the radio emissions from the pulsar and from the companion. In most cases, we will see only the pulsar signal, even if it possesses a companion, because the emission induced by the companion will never be directed toward Earth. This is why one has detected the presence of orbiting planets via pulsar's radio emission timing without necessarily detecting the planet-induced signal. Conversely in the case of FRBs, we see only the emission induced by the companion, because the pulsar's emission is never directed toward Earth (or simply because it is too weak to be detected at extragalactic distances). 

But in the cases where the magnetic axis (or the companion's orbital plane) is tilted enough relative to the pulsar's rotational axis, we may be able to detect both signals, superimposed to each other. Because the companion's signal is very rare (in time) compared to the pulsed signal from the neutron star, it may remain unnoticed except if explicitly searched for, and it will generally be invisible in the folded pulsar signal (at the pulsar period) except if it is very strong. Note also that as pulsar timing is based on the variations of distance from the neutron star to the observer, it is mostly efficient for distant companions, thus a low-mass companion orbiting close to the neutron star may well remain undetected by time of arrival analysis.

\subsection{Observational tests}

The theories that have been previously proposed to explain the Lorimer burst and FRBs predict a unique event from a given source, or irregularly repeatable pulses. However the latter are not consistent with the large observed dispersion measures. On the contrary, our explanation based on highly collimated emission from a pulsar companion's wind wake implies that these pulses should be periodically observable, at the orbital period of the companion. A decisive test of our theory would thus consist in the redetection of a signal with similar characteristics coming from the direction of the source of the Lorimer burst, or of any known FRB, or from PSR J1928+15. We recommend to observe again the same regions of the sky. The difficulty is to observe continuously in order not to ``miss" a pulse with duration of a few msec that may occur at any time within hours, days or more.

Another interesting target is the binary pulsar PSR J2222-0137, that contains a millisecond pulsar orbited by a companion of unknown nature \citep{Deller_2013}. As the system is seen nearly edge-on ($\sin i = 0.9985 \pm 0.0005$) and its orbital parameters are known, it is possible to observe it in search for the companion-induced signal around the transit time of the companion between the pulsar and Earth, or even to search this signal in existing data.

Note that at frequencies about 1 GHz, one expects an angular wandering of the source of the order of $10^{-3}$ arcsec and an angular broadening one order of magnitude larger \citep{Cordes_1990}. Both effects are much smaller than $1/\gamma$ and should not significantly influence the above discussion. However, intensity fluctuations due to strong scintillation may amplify or attenuate pulses, with an amplitude of fluctuation from one pulse to the other up to 100\%. Thus the detection of the Lorimer burst or of FRBs could have benefited from both a perfect alignment and a constructive scintillation.

Other possible tests of our theory related to the frequency range of the signal discussed in section \ref{section_frequencies} includes (i) the existence of periodic (in frequency) intensity fluctuations along the pulse, that would be due to emission at harmonics of a given frequency (the fundamental being the frequency interval between bright spots), (ii) the existence of a low-frequency cutoff of the emission below the fundamental frequency predicted in section \ref{section_frequencies}, and (iii) the polarization of the radio signal: it should be strongly circular or elliptical when produced by the CMI \citep{Zarka_2004a}.

\subsection{Radio waves from pulsar companions as a probe of pulsar winds}

Up to now, research on very low-mass pulsar companions has led to the discovery of five planets \citep{Wolszczan_1992, Thorsett_1993, Bailes_2011} and speculations about orbiting asteroids \citep{Cordes_2008, Deneva_2009,Shannon_2013,Mottez_2013b}. These objects raise a moderate interest in the field of planetology: they orbit stars which left the main sequence a long time ago (in a state that the Sun will never reach), they are exposed to the neutron star X-rays that probably prevent life to develop, they are embedded in a relativistic wind, and they are very far from us. 

By contrast, the study of these objects is of high interest for the understanding and probing of pulsar winds. If companion-induced radio signals are confirmed, then the underlying theory of Alfv\'en wings of bodies in a pulsar wind \citep{Mottez_2011_AWO,Mottez_2011_AWW,mottez_2012c} would be reinforced too. This theory supposes a wind velocity lower than the Alfv\'en wave velocity (both being very close to $c$). This hypothesis, compatible with many theoretical model, is valid only if the pulsar wind at the companion distance is still Poynting flux dominated. In spite of many discussions (see \citet{Bucciantini_2014}
for review) there is no convincing proof of this fact. Observational testing of our theory may bring this missing proof and provide unique measurements of the Lorentz factor of the wind at AU distances, a key ingredient to the theories of dissipation in pulsar winds. The distance range that could be probed is from 10-100 light cylinder radii (otherwise, the companion is destroyed by tidal effects) to a few AU (much closer than the wind termination shock, otherwise at thousands of AU the companion is practically unbound gravitationally to the neutron star). 

If we find that the radio transients that we discussed in section \ref{possible_explanation} are indeed periodic, the shortest measured period will be the orbital period of the companion $T_{orb}$. As neutron stars have a narrow range of masses ($1-3$ $M_\odot$), we can estimate the companion distance to the star. Then, measured pulse characteristics  will allow us to deduce the value of the pulsar wind Lorentz factor $\gamma$  and to know if the wind is Poynting flux dominated at the orbital distance of the companion, i.e. far from the light cyclinder, and far from the termination shock.

\appendix
\section{Orientation of the Alfv\'en wings relative to the local magnetic field} \label{sec_calcul_angle_AW}

The computation of the Alfv\'en waves electric current involves a first invariant $\vec{V}_s$, the vector defining the Alfv\'en characteristics, whose modulus is the speed of an Alfv\'enic perturbation along this characteristics. With $s=\pm 1$ characterizing the two wings, and $\vec{v}$ and $\vec{B}$ the perturbed flow velocity and magnetic field:
\begin{equation} \label{invardansR0} 
\vec{V}_s =\vec{v}-\frac{s \vec{B}(1 - {\vec{v}_0 \cdot \vec{v}}/{c^2})}{ (\lambda -s \vec{B}_0 \cdot \vec{v}/c^2)}=\vec{v}_0-\frac{s \vec{B}_0}{\gamma_0^2 \Lambda}.
\end{equation}
The  last term is an estimate of the invariant involving only unperturbed wind parameters.
The parameter $\lambda$ can be expressed as a function of the unperturbed wind parameters:
\begin{equation} \label{eq_definition_lambda}
\lambda = \left[\mu_0 \rho_0'+c^{-2} {B}_{0}^{r2}+ c^{-2} \gamma_0^{-2} B_0^{ \phi 2} \right] ^{1/2}
\end{equation}
where  $\rho_0'$ is the density in the moving frame of the wind (primed quantities refer hereafter to that frame), and
\begin{equation}
\Lambda = \lambda -s {B}_{0}^r  {v}_0/c^2 \sim \lambda -s {B}_{0}^r/c.
\end{equation}
The invariant $\vec{V}_s$, related to the theory of uncompressible Alfv\'en waves, was derived in \citet{Mottez_2011_AWW} following a linear approximation for the derivation of the general properties of relativistic Alfv\'en wings. This invariant was found again in \citep{mottez_2012c} in the context of the nonlinear theory of relativistic simple Alfv\'en waves. 

Because the direction of the Alfv\'en wings is those of $\vec{V}_s$ and because $\vec{v}_0=v_0^r$,
the angle $\delta$ between the Alfv\'en wing and the radial direction is 
\begin{equation}\label{eq_delta_0}
\delta_s = s\arctan ({B}_0/{ c\Lambda \gamma_0^2})
\end{equation}
When $\Lambda$ is expressed as a function of the wind invariants, we find 
\begin{equation} \label{eq_delta_complet}
\delta_s=s\arctan{ \left[  \frac{x}{\gamma_0 \left(\sqrt{1+x^2(1+\frac{\gamma_0}{\sigma_0})}-s \right)}\right]},
\end{equation}
where $\sigma_0$ is the magnetization factor defined by
\begin{equation}
\sigma_0 = \frac{\Omega_*^2 \Psi^2 }{\mu_0 f c^3}.
\end{equation}
The MHD models considering a relativistic radial wind \citep{Arons_2004,Kirk_2009} imply an asymptotic Lorentz factor scaling as 
\begin{equation} \label{eq_gamma_infty}
\sigma_0 \sim \gamma_0^3 
\end{equation}
and the term ${\gamma_0}/{\sigma_0}$ can be neglected. Actually, it is negligible as long as the wind is Poynting flux dominated, that is 
the basic assumption that govern this paper. 
When ${\gamma_0}/{\sigma_0}$ is neglected, we obtain Eq. (\ref{eq_delta}).

%%%%%%%%%%%%%%%%%%%%%%%%%%%%%%%%%%%%%%%%%%%%%%%%%
%%%%%%%%%%%%%%%%%%%%%%%%%%%%%%%%%%%%%%%%%%%%%%%%%
%%%%%%%%%%%% T A B L E S %%%%%%%%%%%%%%%%%%%%%%
%%%%%%%%%%%%%%%%%%%%%%%%%%%%%%%%%%%%%%%%%%%%%%%%%
%%%%%%%%%%%%%%%%%%%%%%%%%%%%%%%%%%%%%%%%%%%%%%%%%

\section*{Acknowlegments}
We thank Michal Bejger (Copernicus Astronomical Center, Polish Academy of Sciences) for suggesting the relevance of pulsar-planet physics to orbiting white dwarfs. This work was partly supported by the CNRS/INSU/PNHE Programme National Hautes Energies. 

%%%%%%%%%%%%%%%%%%%%%%%%%%%%%%%%%%%%%%%%%%%%%%%%%
%%%%%%%%%%%%%%%%%%%%%%%%%%%%%%%%%%%%%%%%%%%%%%%%%
%%%%%%%%%%%% REFERENCES %%%%%%%%%%%%%%%%%%%%%%
%%%%%%%%%%%%%%%%%%%%%%%%%%%%%%%%%%%%%%%%%%%%%%%%%
%%%%%%%%%%%%%%%%%%%%%%%%%%%%%%%%%%%%%%%%%%%%%%%%%

%\end{thebibliography}{}
\bibliographystyle{aa} % style aa.bst
%\bibliography{aile_alfven} % your references Yourfile.bib
\bibliography{article} % your references Yourfile.bib

%%%%%%%%%%%%%%%%%%%%%%%%%%%%%%%%%%%%%%%%%%%%%%%%%
%%%%%%%%%%%%%%%%%%%%%%%%%%%%%%%%%%%%%%%%%%%%%%%%%
%%%%%%%%%%% F I G U R E S %%%%%%%%%%%%%%%%%%%%%
%%%%%%%%%%%%%%%%%%%%%%%%%%%%%%%%%%%%%%%%%%%%%%%%%
%%%%%%%%%%%%%%%%%%%%%%%%%%%%%%%%%%%%%%%%%%%%%%%%%

\end{document}